\documentclass[10pt,letterpaper,compressed]{article}

\usepackage{opex3}

\begin{document}

\title{Suppression of coherence degradation due to plasma contribution in high-power supercontinuum
generation in coated metallic hollow waveguides}
\author{A. Husakou$^*$ and J. Herrmann}

\address{Max Born Institute for Nonlinear Optics and Short Pulse
Spectroscopy\\ Max Born Str. 2a, D-12489 Berlin, Germany}
\email{gusakov@mbi-berlin.de}

\begin{abstract}
We study  the suppression of noise
degradation in high-power soliton-induced supercontinuum generation under the influence of plasma contributions in metallic dielectric-coated hollow
waveguides. The high coherence of the supercontinuum is related to the coherent seed components formed by the abruptly rising plasma density. We predict the generation of highly coherent supercontinua with two-octave broad spectra and spectral power densities in the range of MW/nm in such waveguides. 
\end{abstract}

\ocis{(030.1640) Coherence; (190.7110) Ultrafast nonlinear optics}

Many applications require light sources which share with a laser
its coherent and uindirectional properties but span a much broader spectral
range like an electric bulb. Such coherent white-light sources
(supercontinuum, SC) was achieved by the application of microstructure fibers
(MF)\cite{1,2}. When a femtosecond pulse with only nJ energy from a laser
oscillator is focused into such fiber, a dramatic conversion from narrowband
light to two-octaves broad spectra is observed \cite{3}.  The SC generation
in MFs in the anomalous dispersion range is connected with the splitting of
the input pulse into several fundamental solitons, which emit phase-matched
non-solitonic radiation \cite{4,5}. The threshold for this highly efficient
mechanism is significantly lower than for self-phase modulation or any other
known spectral broadening process. Therefore soliton-induced supercontinuum
generation has attracted significant interest and found many important
applications ranging from frequency metrology and optical
coherence tomography to spectroscopy, confocal microscopy, medicine and others (for an overview see \cite{6}). 

For many applications the noise and coherence properties of this
octave-spanning white light is of crucial importance. In particular, for pulse compression a shaper is needed to adjust the phases of the different spectral components with respect to each other. Therefore the
pulse-to-pulse phase noise of the spectral components is very detrimental for pulse compression. Low-noise supercontinua are also essential for optical
frequency metrology, optical coherence tomography, and other
applications. Several experimental studies and theoretical simulations have
shown that in general the SC coherence is very sensitive to both the
fundamental quantum noise and the shot-to-shot pump intensity
fluctuations (technical noise) [7-14]
 These studies have shown that a
sufficient level of coherence of the SCs can only be obtained using pulses
with a relatively small intensity and durations of about 50 fs or less \cite{8,12,14}, or for input wavelengths
tuned deeper into the anomalous dispersion region\cite{13}.

\begin{figure}[!ht]
 \centering\includegraphics[width=0.6\textwidth,clip]{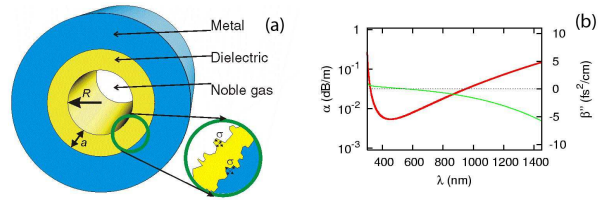}%
\caption{Scheme (a) and the loss (red curve) and GVD (green curve) in (b) for a silver waveguide with diameter $D=80$ $\mu$m and fused-silica coating thickness $a$ of 45 nm. The roughness size is indicated by $\sigma$.
\label{sch}}
\end{figure}

A further disadvantage of SC generation in MFs is the restricted
maximum peak power densities of SCs limited to a range of tens of W/nm due to the small radii in MFs and the material damage of the solid core.
Recently we predicted that a significant increase of the peak power density by up to five orders of magnitude is possible by the use of specific noble-gas-filled hollow waveguides a dielectric-coated metal cladding, which can be produced by chemical vapor deposition\cite{19,20}. In Fig. \ref{sch}(a), the geometry of such waveguide is presented. The hollow core of the fibre with diameter $D$ is surrounded by a metallic cladding (blue) coated on the inner side by a dielectric material such as fused silica (yellow) with thickness $a$.

In the present paper we show that, besides its very high spectral peak power density, soliton-induced SCs in metallic hollow waveguides coated with a dielectic has the additional advantage of suppressed coherence degradation due to the contribution of the plasma.

For the numerical simulations we use a model based on a generalization of
the unidirectional equation of Ref.\cite{4}  with inclusion of higher-order
transverse modes and the effects of the ionization and plasma formation in
the waveguide.
The Fourier transform $\vec{E}(x,y,z,\omega)=\sum_jF_j(r,\omega)E_j(z,\omega)$ of the electromagnetic field $\vec{E}(z,x,y,t)$ in
the waveguide can be decomposed into transverse modes $F_{j}(r,\omega)$,
where $j$ is the mode number,
with $z$ being the propagation coordinate and $r^2=x^2+y^2$. We consider only linearly-polarized input fields exciting
azimuthally-symmetric modes EH$_{1j}$ with the same polarization. Substituting this ansatz into the wave equation and neglecting
weak backreflected field components the following system of first-order
differential equations can be derived
\begin{eqnarray}
\frac{\partial E_{j}(z,\omega )}{\partial z}=i\beta _{j}(\omega
)E_{j}(z,\omega )-\frac{\alpha _{j}}{2}(\omega )E_{j}(z,\omega
)\nonumber\\+\frac{i\omega^2}{2c^2\epsilon_0\beta_j(\omega)}P_{NL}^{(j)}(z,\omega )
\end{eqnarray} 
where $\beta _{j}(\omega )$ and $\alpha _{j}(\omega )$ are the wavenumber
and the loss, $F_{j}(r,\omega )$ the transverse mode shape and $
P_{NL}^{(j)}(z,\omega )$ is the nonlinear polarization of the mode $j$.

Note that this approach does not use the slowly-varying-envelope
approximation which allows to describe adequately the generation of
ultrabroadband spectra. 
The inclusion of higher-order transverse modes takes
into account a possible
 energy transfer to higher-order modes by the nonlinear polarization. $%
\beta _{j}(\omega )$, $\alpha _{j}(\omega )$ and $F_{j}(r,\omega )$, are
calculated by the transfer-matrix approach assuming a circular
waveguide structure as shown in Fig. \ref{sch}(a) of the paper. Additionally, roughness
loss is included in the calculation of $\alpha _{j}(\omega )$ using the
model of pointlike scatterers; for further details of the transfer matrix
theory and roughness loss calculation, see Ref. \cite{22}.

The Fourier transform $P_{NL}^{(j)}(z,\omega )$ of the nonlinear
polarization for the transverse mode $j$ is given by
\begin{equation}
P_{NL}^{(j)}(z,\omega )=\int_{0}^{R}\!\!2\pi rF_{j}(r,\omega
)\int_{-\infty }^{\infty }\!\!\exp (i\omega t)P_{NL}(z,r,t)drdt
\end{equation}%
where $F_j(r,\omega)$ is the mode profile of the j-th mode.
The nonlinear polarization $P_{NL}(z,r,t)$ includes three terms: the Kerr
nonlinearity, the plasma-induced refraction index change and absorption due
to ionization:
\begin{eqnarray}
P_{NL}(z,r,t)=\epsilon _{0}\chi _{3}E^{3}(z,r,t)-\rho
(z,r,t)ed(z,r,t)-\nonumber\\{E_{g}\epsilon _{0}c}\int_{-\infty}^t\frac{E(z,r,t)}{\tilde{I}(z,r,t)}\frac{\partial \rho (z,r,t)%
}{\partial t}dt
\end{eqnarray}%
where $\chi_{3}=(4/3)c\epsilon_0n_2$ is the third-order
polarizability of the gas filling and $\tilde{I}(z,r,t)$ is the intensity averaged over few optical periods.
For argon we have $n_2=1\times10^{-19}$ cm$^2$/W at 1 atm, $E_{g}$
is the ionization potential, $\rho (z,r,t)$ is the electron density
and $d(z,r,t)$ is the mean free-electron displacement in the
electric field. The evolution of the two latter quantities is
described by
\begin{equation}
\frac{\partial \rho (z,r,t)}{\partial t}=(N_0-\rho (z,r,t))\Gamma (\tilde{I}%
(z,r,t))
\end{equation}%
\begin{equation}
\frac{\partial^2 d(z,r,t)}{\partial t^2}=-eE(z,r,t)/m_{e}-\frac{\partial_td(z,r,t)}{\rho (z,r,t)}%
\frac{\partial \rho (z,r,t)}{\partial t}
\end{equation}%
where $\Gamma (\tilde{I}(z,r,t))$ is the photoionization rate
calculated from the Keldysh-Faisal-Reiss model\cite{frk} which describes both
the multiphoton and tunnelling regime and $N_0$ is the initial argon density.
In the following we study the coherence degradation of the supercontinua due to the influence of the fundamental quantum noise. In the Wigner quasi-probability representation the evolution equations for quatum field operators are mapped onto the classical equations \cite{16} where the effect of
quatum noise is included by adding to the input field $E(t)$ the quantum shot
noise $\Delta E(t)$ defined by $<\Delta E(t_{1})\Delta E(t_{2})>=\delta
(t_{1}-t_{2})\hbar \omega _{0}/(2E_{tot})$ where $\omega _{0}$ is the input
frequency, $E_{tot}$ is the total energy of the pulse and the average $<...>$ is taken over the noise realisations (see Ref. \cite{16}
for details). The quantity which characterizes the coherence of the SC is
given by the first order coherence function $g(\lambda )$ which directly
corresponds to the visibility measured in interference experiments and is
defined as 
\begin{equation}
g(\lambda )=\Re \left[ \frac{<E_{b}(\lambda )E_{a}^{\ast }(\lambda
)>_{ab,a\neq b}}{<E_{a}(\lambda )E_{a}^{\ast }(\lambda )>_{a}}\right] 
\end{equation}%
where $E_{a}(\lambda )$ is the spectrum in the fundamental mode after the waveguide and the indices $a,b$ denote the realization.

\begin{figure}[!ht]
\centering\includegraphics[width=0.85\textwidth]{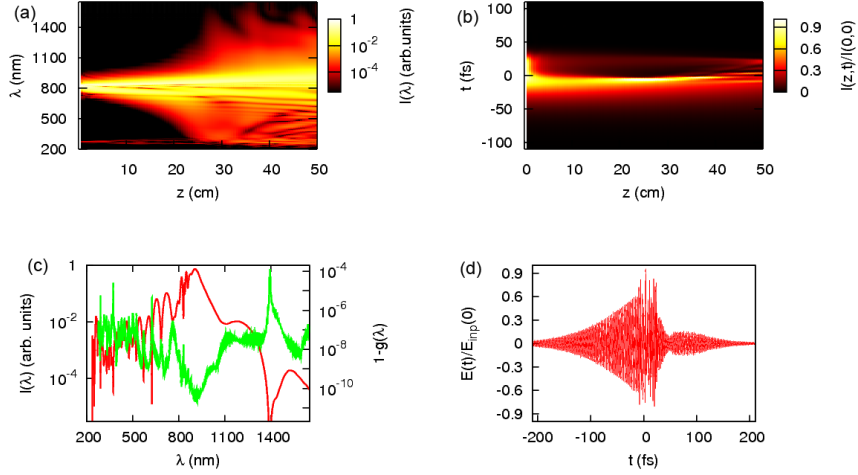}%
\caption{Evolution of the spectrum (a) and temporal shape (b) for the propagation of a 100-TW/cm$^2$, 100-fs pulse in a fused-silica-coated silver waveguide. In (c) the output spectrum (red) and incoherence $1-g(\lambda)$ (green) are shown and in (d) the ouput temporal shape are presented. The propagation length is 50 cm, input wavelength is 800 nm, the waveguide parameters are $D=80$ $\mu$m, $a=45$ nm.
\label{HE}}
\end{figure}

We consider a hollow waveguide with a diameter of 80 $\mu$m, which is  made
from a silver cladding coated from the inner side by 45-nm-thick layer of
fused silica [ Fig. \ref{sch}(a)].
The  loss and dispersion properties of the waveguide are illustrated in Fig. \ref{sch}(b). It can be seen in Fig. \ref{sch}(b)
 that the loss remains
relatively low in the range of 10$^{-2}$-10$^{-1}$ dB/m for wavelengths from 0.4 $\mu$m to 1.4 $\mu$m.
This means that at most 10\% of the energy is lost during a propagation over 1 m, while
in a conventional dielectric hollow waveguide with the same diameter, roughly 80\% of the
energy is lost. The physical reason for this difference is that the
reflection coefficient for grazing incidence of light is higher for
fused-silica-coated silver than for a layer of fused silica. Therefore the diameter of the waveguides can be reduced, which leads to a significant modification of the dispersion properties and an extended range of anomalous dispersion\cite{22}. The group velocity dispersion, illustrated in
Fig. \ref{sch}(b) by the green curve, is anomalous for $\lambda >570$ nm at 1 atm of the argon filling.

     We now study the evolution of a 100-fs input pulse at 830 nm with a peak intensity of 100 TW/cm$^2$
propagating a distance of 50 cm in the above-described fiber. 
It can
be seen in Fig. \ref{HE}(a),(c) that the generated radiation after 50 cm of propagation covers the spectral
range from 200 nm to 1400 nm, with a total width corresponding to
more than two octaves. In Fig. \ref{HE}(b) the evolution of the temporal shape of the pulse is illustrated, which shows pulse compression and formation of an edge in the middle of the pulse with sharply decreasing intensity after the edge. During further propagation distances isolated temporal peaks start to form, as shown in Fig. \ref{HE}(d). These peaks can be identified as
fundamental solitons caused by the anomalous dispersion at the input
wavelength together with the Kerr nonlinearity of the argon filling. Similar to the case of soliton-induced SC generation in MFs \cite{4}
the two-octave broad spectrum represented by the red curve in Fig. \ref{HE}(c) comprise the spectrum of several solitons red-shifted by the
recoil effect and at the short-wavelengh
side the non-solitonic radiation emitted by them. The wavelength-averaged peak power spectral density is about 10$^6$
W/nm with an intensity at the output of about 10$^{14}$ W/cm$^2$. In Fig. \ref{HE}(c) the quantity $1-g(\lambda)$ (first-order incoherence) is
shown by the green curve demonstrating a surprisingly high coherence which deviates from unity by no more than 10$^{-5}$.
This high coherence of soliton-induced SC
generation in hollow waveguides is in contrast to SC generation in MFs based on the same broadening mechanism.

\begin{figure}[!ht]
\centering\includegraphics[width=0.8\textwidth]{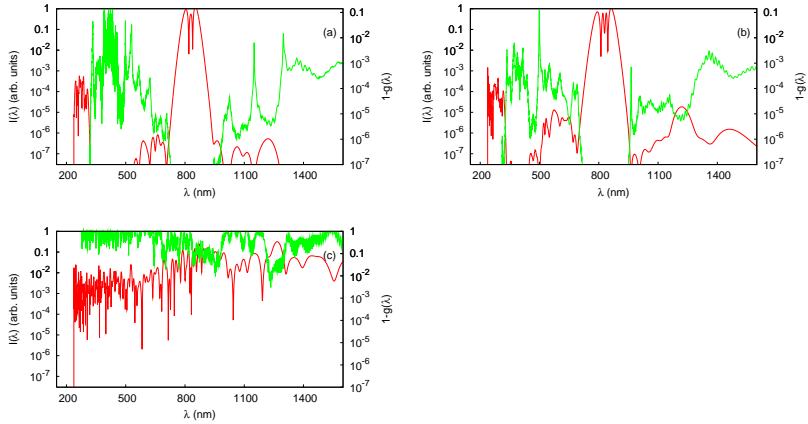}%
\caption{Spectrum and coherence for a propagation model without plasma contribution. The spectrum (red curve) and coherence (green curve) are presented at the distances of 15 cm (a), 20 cm (b) and 50 cm (c). The input pulse and waveguide parameters are the same as in Fig. \ref{HE}.
\label{enp}}
\end{figure}

\begin{figure}[!ht]
\centering\includegraphics[width=0.8\textwidth]{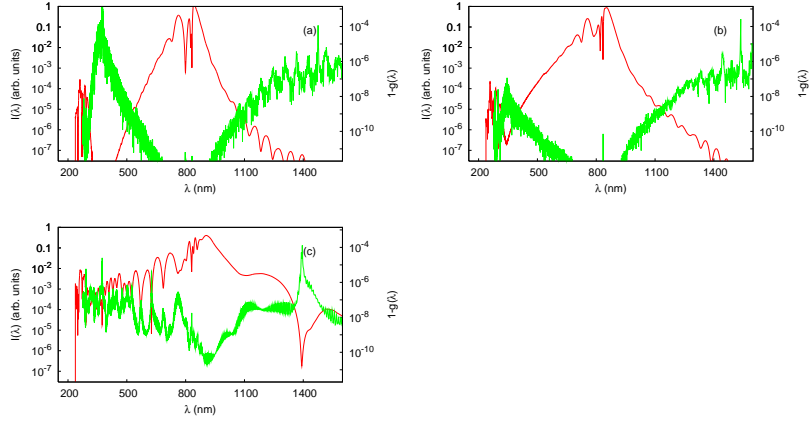}
\caption{Spectrum and coherence for a propagation model with plasma contribution. The spectrum (red curve) and coherence (green curve) are presented at the distances of 15 cm (a), 20 cm (b) and 50 cm (c). The input pulse and waveguide parameters are the same as in Fig. \ref{HE}.
\label{ep}}
\end{figure}

To understand the suppression of coherence
degradation for soliton-induced SC generation in metallic hollow waveguide we study the role of the plasma
contribution in the SC formation process.
In Figs. \ref{enp} and \ref{ep} we present the evolution of the spectrum and incoherence function $1-g(\lambda)$ for models without (Fig. \ref{enp}) and with (Fig. \ref{ep}) plasma contributions. It can be seen that without plasma, at the propagation length of 15 and 20 cm   seed spectral components are formed [Fig. \ref{enp}(a) and (b)] within spectral wings between 500 nm and 700 nm as well as around 1100 nm. They arise by the
amplification of noise due to the gain of the modulation instability. The coherence of the side maxima decreases with propagation from 15 to 30 cm, with an incoherence on the level of 10$^{-4}$-10$^{-3}$.  With further propagation after fission of the pulse into several solitons,
the spectrum is dramtically broadened as shown by the red curve in Fig. \ref{enp}(c), but the coherence is also significantly degraded (green curve) with a low  average coherence of the output spectrum $\widetilde{g(\lambda)}=0.63$.  In contrast, in the case when the plasma contribution is included in the numerical model (Fig. \ref{ep}), the seed components arise mainly on the short-wavelength side of the spectrum, forming a smooth spectral wing from 300 to 700 nm in Fig. \ref{ep}(b). The mechanism of this smooth short-wavelength wing formation is related to the abrupt rise of the plasma density near the peak of the pulse  leading to a strong phase modulation due to time-dependent plasma contribution in the refractive index. We note that this wing is highly coherent in our case with $1-g(\lambda)\sim 10^{-9}$ [green curve in Fig. \ref{ep}(b)], as should be expected since its spectral phase is determined by the phases of the input pulse, similar to the case of self-phase modulation. 
Now in contrast with Fig. \ref{enp}(c)  further propagation and spectral
broadening of the pulse up to 50 cm do not lead to a dramatic coherence
degradation but to the preservation of the high coherence of the SC. The
resulting spectrum [Fig. \ref{ep}(c) and Fig. \ref{HE}(c)] is characterised
by the average coherence of $\widetilde{g(\lambda)}=1-3\mathrm{x}10^{-6}$. 

\begin{figure}[!ht]
\centering\includegraphics[width=0.75\textwidth]{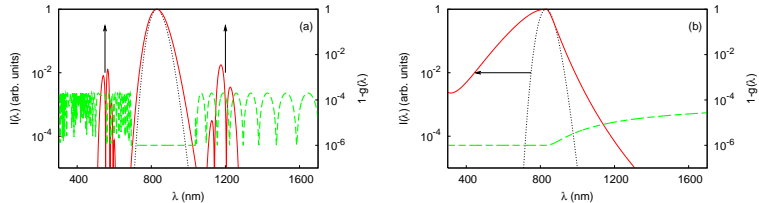}%
\caption{Schematic presentation of the supercontinuum generation at the initial stage. Black dotted curve indicates the input spectrum, the red curve is the spectrum at the position of the seed component formation, green dashed curve is the incoherence $1-g(\lambda)$. In  (a),arrows indicate the four-wave-mixing gain bands; in (b), the arrow indicates the spectral broadening induced by plasma.
\label{exp}}
\end{figure}

The mechanism of the suppression of coherence degradation in
soliton-induced SC generation is schematically presented in Fig. \ref{exp}(a,b). In the case of a relatively long and intense pulse due to gain by modulation instability seed spectral components arise {\it from noise}, as represented in Fig. \ref{exp}(a) by the black arrows. These seed components are further amplified by the gain produced by the fundamental solitons after fission due to third-order dispersion leading to the large spectral broadening but accompanied by coherence degradation. In contrast,
in the case of the seed components formation due to plasma, as schematically presented in Fig. \ref{exp}(b) the short-wavelength wing arise {\it from the pulse itself}. It is smooth and highly coherent, and after amplification by the gain which leads to the emission of non-solitonic radiation the coherence remains high.

\begin{figure}[!ht]
\centering\includegraphics[width=0.45\textwidth]{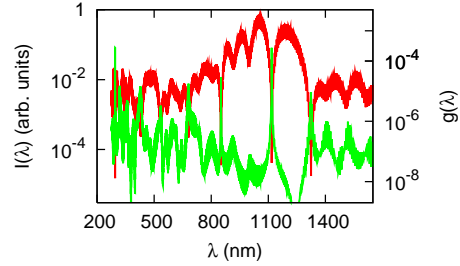}%
\caption{Output spectrum (red) and coherence (green) for the propagation of a 100-TW/cm$^2$, 10-fs pulse in a fused-silica-coated silver waveguide. The propagation length is 50 cm, input wavelength is 800 nm, the waveguide parameters are $D=80$ $\mu$m, $a=45$ nm.
\label{sh}}
\end{figure}

For comparison we calculated the spectrum and coherence function for a shorter input pulse, as illustrated in Fig. \ref{sh}.
The spectrum in this case is not significantly broader although slighly smoother, and the coherence increases to values in the range of $1-10^{-7}$. In this case the input spectrum is already quite broad, which facilitates the formation of the coherent sidebands from the broad spectrum during the soliton compression and leads to high final coherence.

\begin{figure}[!h]
\centering\includegraphics[width=0.75\textwidth]{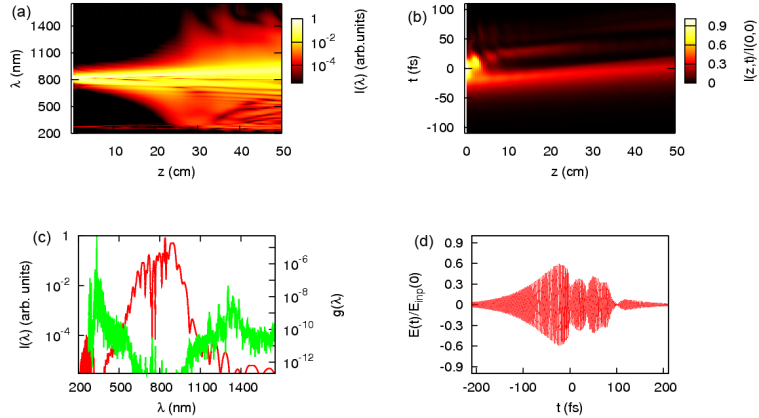}%
\caption{Evolution of the spectrum (a) and temporal shape (b) for the propagation of a 100-TW/cm$^2$, 100-fs pulse in a fused-silica-coated silver waveguide. In (c) the output spectrum (red) and the coherence (green) are shown and in (d) the ouput temporal shape are presented. The propagation length is 50 cm, input wavelength is 800 nm, the waveguide parameters are $D=200$ $\mu$m, $a=45$ nm.
\label{norm}}
\end{figure}
Finally, we have also performed simulations for a larger waveguide diameter of 200 $\mu$m with smaller waveguide contribution to dispersion and normal GVD at the input wavelength. The results are presented in Fig. \ref{norm}. In the temporal shape shown in Fig. \ref{norm}(b),(d) a modulation is formed but without sharp edge, and the spectrum shown in Fig. \ref{norm}(a) and (c) broadens somewhat but remains relatively narrow. This indicates that the anomalous dispersion and the soliton mechanism are essential for the predicted two-octave spectral broadening in the considered hollow waveguide, which can not be achieved by
self-phase modulation and plasma-related broadening. The coherence shown by the green curve in Fig. \ref{norm}(c) is very high, typical for spectral broadening by SPM.

In conclusion, we studied the generation of high-power soliton-induced supercontinua in the anomalous dispersion range of 
metallic dielectric-coated hollow waveguides filled with argon under the influence of plasma contributions to the spectral broadening process. We predicted that the coherence degradation in soliton-induced SC generation as typically observed in MFs can be suppressed by the plasma contributions. The coherence preservation is explained by the formation of seed components of the SC generation due to sharp raise of the plasma density. In this way highly coherent supercontinua with two-octave broad spectra and spectral power densities in the range of MW/nm can be achieved in such waveguides. These findings could have applications in a wide
range of fields in which coherent high-power supercontinua are required.

\end{document}